\documentclass[]{siamart}

\usepackage{lipsum}
\usepackage{graphicx}
\usepackage{subfig}
\usepackage{amsfonts}
\usepackage{epstopdf}
\usepackage{booktabs}
\usepackage{bm}
\usepackage{psfrag}
\usepackage[utf8]{inputenc}
\usepackage{algorithm,algorithmic,amssymb,amsmath,letltxmacro}
\newlength{\commentindent}
\makeatletter
\renewcommand{\algorithmiccomment}[1]{\unskip\hfill\makebox[\commentindent][l]{//~#1}\par}
\LetLtxMacro{\oldalgorithmic}{\algorithmic}
\renewcommand{\algorithmic}[1][0]{%
  \oldalgorithmic[#1]%
  \renewcommand{\ALC@com}[1]{%
    \ifnum\pdfstrcmp{##1}{default}=0\else\algorithmiccomment{##1}\fi}%
}
\makeatother

\ifpdf
  \DeclareGraphicsExtensions{.eps,.pdf,.png,.jpg}
\else
  \DeclareGraphicsExtensions{.eps}
\fi

\newcommand{\TheTitle}{Landau Collision Integral Solver with Adaptive Mesh Refinement on Emerging Architectures} 
\newcommand{\HTitle}{Landau Collision Integral with AMR on emerging architectures} 
\newcommand{\TheAuthors}{M. F. Adams, E. Hirvijoki, M. G. Knepley, J. Brown, T. Isaac, and R. Mills}

\headers{\HTitle}{\TheAuthors}

\title{{\TheTitle}}

\author{
  Mark F. Adams\thanks{Scalable Solvers Group, Lawrence Berkeley Laboratory, Berkeley CA (\email{mfadams@lbl.gov})}
 \and
 Eero Hirvijoki \thanks{Princeton Plasma Physics Laboratory, Princeton, NJ, USA}
 \and
  Matthew G. Knepley\thanks{Computational and Applied Mathematics, Rice University}
 \and
 Jed Brown\thanks{Department of Computer Science, University of Colorado, Boulder}
  \and 
  Tobin Isaac \thanks{Computational and Applied Mathematics, Rice University}
  \and 
  Richard Mills \thanks{Intel Corporation}
}

\usepackage{amsopn}

\newcommand{\Order}[1]{\ensuremath{\mathcal{O}(#1)}}    

\newcommand{\appropto}{\mathrel{\vcenter{
  \offinterlineskip\halign{\hfil$##$\cr
    \propto\cr\noalign{\kern2pt}\sim\cr\noalign{\kern-2pt}}}}}
    
\begin{document}

\date{}

\maketitle

\begin{abstract}
  The Landau collision integral is an accurate model for the
  small-angle dominated Coulomb collisions in fusion plasmas.  We
  investigate a high order accurate, fully conservative, finite
  element discretization of the nonlinear multi-species Landau
  integral with adaptive mesh refinement using the PETSc library
  (www.mcs.anl.gov/petsc).  We develop algorithms and techniques to
  efficiently utilize emerging architectures with an approach that
  minimizes memory usage and movement and is suitable for vector
  processing.  The Landau collision integral is vectorized with Intel
  AVX-512 intrinsics and the solver sustains as much as 22\% of the
  theoretical peak flop rate of the Second Generation Intel Xeon Phi,
  Knights Landing, processor.
\end{abstract}

\begin{keywords}
  Landau collision integral, fusion plasma physics
\end{keywords}

\section{Introduction}
  
The simulation of magnetized plasmas is of commercial and scientific
interest and is integral to the DOE's fusion energy research program
\cite{Exascale,ISW,TW}.
Although fluid models are widely employed to model fusion plasmas, the
weak collisionality and highly non-Maxwellian velocity distributions in such plasmas motivate
the use of kinetic models, such as the so-called {\it Vlasov-Maxwell-Landau} system. The evolution of the phase-space density or
distribution function $f$ of each species (electrons and multiple
species of ions in general) is modeled with
\begin{equation*}
\frac{df}{dt}\equiv
\frac{\partial f}{\partial t} + \frac{\partial\bm x}{\partial t}
\cdot \nabla_x f+ \frac{\partial\bm v}{\partial t} \cdot \nabla_v f
= \frac{\partial f}{\partial t} + {\bm v} \cdot \nabla_x f+
\frac{e}{m}\left( {\bm E} + {\bm v} \times {\bm B} \right) \cdot
\nabla_v f = C
\end{equation*}
where $e$ is charge, $m$ mass, ${\bm E} $ electric field, ${\bm B}$
magnetic field, ${\bm x}$ spatial coordinate, ${\bm v}$ velocity
coordinate, and $t$ time. The Vlasov operator $d/dt$ describes the
streaming of particles influenced by electromagnetic forces, the
Maxwell's equations provide the electromagnetic fields, and the Landau
collision integral~\cite{landau1936kinetic}, $C$, dissipates entropy and embodies the
transition from many-body dynamics to single particle statistics. As
such, the Vlasov-Maxwell-Landau system-of-equations is the gold
standard for high-fidelity fusion plasma simulations. 


The Vlasov-Maxwell-Landau system also conserves energy and momentum,
and guaranteeing these properties in numerical simulations is critical
to avoid plasma self heating and false momentum transfer during long-time simulations. Hirvijoki and Adams
recently developed a finite element discretization of the Landau
integral, which is able to preserve the conservation properties of the
Landau collision integral with sufficient order accurate finite
element space \cite{Hirvijoki2016}.  We now continue this work with
the development of a multi-species Landau solver with adaptive mesh
refinement (AMR), which is designed for emerging architectures and implemented on the Second Generation Intel Xeon Phi,
Knights Landing (KNL), processor.

Due to the nonlinearity of the Landau collision integral, it has an intensive work complexity
of \Order{N^2} with $N$ global integration or quadrature points.
Given this high-order work complexity, reducing the total number of
quadrature points decreases computational cost substantially.  We use
high order accurate finite elements and AMR to maximize the
information content of each quadrature point and thus minimize the
solver cost.  We adapt nonconforming tensor product meshes using the
{\it p4est} library
\cite{Stadler1033,DBLP:journals/siamsc/IsaacBWG15,Rudi:2015:EIS:2807591.2807675},
as a third party library in the PETSc library
\cite{petsc-web-page,petsc-user-ref}.

We develop algorithms and techniques for optimizing the Landau solver
on emerging architectures, with emphasis on KNL, and verify the solver
on a model problem.  We vectorize the kernel using Intel AVX-512
intrinsics and achieve a flop rate as high as $22\%$ of the
theoretical peak floating point rate of KNL.

\section{Conservative Finite Element Discretization of the Landau Integral}
\label{sec:landaufem}

We consider the multi-species version of the conservative finite
element discretization of the Landau collision integral presented by
Hirvijoki and Adams \cite{Hirvijoki2016}.  Under small-angle dominated
Coulomb collision, the distribution function $f_{\alpha}(\bm{v},t)$ of
species $\alpha$ evolves according to
\begin{equation}
\label{eq:landau}
\frac{\partial f_{\alpha}}{\partial t}= \sum_{\beta}\nu_{\alpha\beta}\frac{m_o}{m_{\alpha}}\nabla_{v} \cdot\int \limits_{\bar\Omega} d\bm{\bar{v}}\;\mathbf{U}(\bm{v},\bm{\bar{v}})\cdot\left(\frac{m_o}{m_{\alpha}}\bar{f}_{\beta}\nabla_{v} f_{\alpha} - \frac{m_o}{m_{\beta}}f_{\alpha} \bar \nabla_{\bar{v}} \bar{f}_{\beta}\right).
\end{equation}
Here
$\nu_{\alpha\beta}=e_{\alpha}^2e_{\beta}^2\ln\Lambda_{\alpha\beta}/(8\pi
m_o^2\varepsilon_0^2)$, $\ln\Lambda$ is the Coulomb logarithm, $m_o$
is an arbitrary reference mass, $\varepsilon_0$ is the vacuum
permittivity, $m$ is mass, $e$ is electric charge, and $\bm{v}$ is the
velocity. Overbar terms are evaluated on the $\bm{\bar{v}}$ grid that covers the domain $\bar\Omega$ of species $\beta$.  The
Landau tensor $\mathbf{U}(\bm{v},\bm{\bar{v}})$ is a scaled projection
matrix defined as
\begin{equation}
\label{eq:landau_tensor}
\mathbf{U}(\bm{v},\bm{\bar{v}})=\frac{1}{\lvert\bm{v}-\bm{\bar{v}}\rvert^3}\left(\lvert\bm{v}-\bm{\bar{v}}\rvert^2\mathbf{I}-(\bm{v}-\bm{\bar{v}})(\bm{v}-\bm{\bar{v}})\right)
\end{equation}
and has an eigenvector $\bm{v}-\bm{\bar{v}}$ corresponding to a zero
eigenvalue.

Given a test function $\psi(\bm{v})$, the weak form of the Landau operator (\ref{eq:landau}) for species $\alpha$ is given by
\begin{equation}
\label{eq:weak-form}
\left(\psi,\frac{\partial f_{\alpha}}{\partial t}\right)_{\Omega}=\sum_{\beta}\left(\psi, f_{\alpha}\right)_{\bm{K},\alpha\beta}+\left(\psi,f_{\alpha}\right)_{\mathbf{D},\alpha\beta}
\end{equation}
where $(\cdot,\cdot)_{\Omega}$ is the standard $L^2$ inner product in $\Omega$ and the weighted inner products present the advective and diffusive parts of the Landau collision integral
\begin{align}
\label{eq:K}
\left(\psi, \phi\right)_{\bm{K},\alpha\beta}&=\int \limits_{\Omega}d\bm{v}\nabla_{v}\psi\cdot \hat{\nu}_{\alpha\beta}\frac{m_o}{m_{\alpha}} \frac{m_o}{m_{\beta}}\bm{K}(f_{\beta},\bm{v})\, \phi,\\
\label{eq:D}
\left(\psi,\phi\right)_{\mathbf{D},\alpha\beta}&=-\int \limits_{\Omega}d\bm{v}\nabla_{v}\psi\cdot\hat{\nu}_{\alpha\beta}\frac{m_o}{m_{\alpha}}\frac{m_o}{m_{\alpha}}\mathbf{D}(f_{\beta},\bm{v})\cdot\nabla_{v}\phi
\end{align}
The collision frequency is normalized with
$\hat{\nu}_{\alpha\beta}=\nu_{\alpha\beta}/\nu_o$ so that time $t$ is dimensionless, and $f_{\beta}$ is
the distribution function of species $\beta$.  The vector
$\bm{K}$ and the tensor $\mathbf{D}$ are defined as
\begin{align}
\label{eq:fric_coef}
  \bm{K}(f,\bm{v})&=\int \limits_{\bar\Omega}d\bm{\bar{v}}\;\mathbf{U}(\bm{v},\bm{\bar{v}})\cdot\bar{\nabla}_{\bar{v}}f(\bm{\bar{v}}), \\
\label{eq:diff_coef}
  \mathbf{D}(f,\bm{v})&=\int
  \limits_{\bar\Omega}d\bm{\bar{v}}\;\mathbf{U}(\bm{v},\bm{\bar{v}})f(\bm{\bar{v}}).
\end{align}
Assuming a finite-dimensional vector space $V_h$ that is spanned by the set of functions $\{\psi_i\}_{i}$, the finite-dimensional approximation of the weak form (\ref{eq:weak-form}) can be written in a matrix form
\begin{equation}
\label{eq:matrix-form}
\mathbf{M}\dot{\bm{f}}_{\alpha}=\mathbf{C}_{\alpha}[\bm{f}]\bm{f}_{\alpha}
\end{equation}
where $\bm{f}_{\alpha}$ is the vector containing the projection coefficients of $f_{\alpha}$ onto $V_h$ and the vector $\bm{f}$ is the collection of all species $\bm{f}_{\alpha}$, The mass and collision matrices are defined
\begin{equation}
\mathbf{M}_{ij}= (\psi_i,\psi_j)_{\Omega},\quad\mathbf{C}_{\alpha,ij}[\bm{f}]=\sum_{\beta=1}^S\left(\psi_i, \psi_j\right)_{\bm{K},\alpha\beta}+\left(\psi_i,\psi_j\right)_{\mathbf{D},\alpha\beta}
\label{eq:fullC}
\end{equation}
The integrals in
(\ref{eq:fric_coef},\ref{eq:diff_coef}), with the Landau tensors in the kernel, have
\Order{N} work for each species $\beta$ and each equation in
(\ref{eq:weak-form}).  With \Order{N} equation this leads to an
\Order{N^2} algorithm for computing a Jacobian or residual when solving the equations (\ref{eq:matrix-form}) for each species.



\section{Algorithm Design for Emerging Architectures}
\label{sec:knlopt}

This section discuses the algorithms and techniques used to
effectively utilize emerging architectures for a Landau integral
solver.  While the Landau operator has \Order{N^2} work complexity,
this work is amenable to vector processing.  We focus on KNL, but the
algorithm is design to minimized data movement and simplify access
patterns, which is beneficial for any emerging architecture. 

The discrete Landau Jacobian matrix construction, or residual
calculation, can be written as six nested loops.  Algorithm
\ref{fig:algo0} shows high level pseudo-code for construction the
Landau Jacobian matrix, with $|G|$ cells in the set $G$, $Nq$
quadrature points in each element, distribution functions $f$, $S$
species, and weights $w_{q_j} = |J\left(q_j\right)| \cdot q_j.weight
\cdot q_j.r$, where $q_j.r$ is the axisymmetric term of the element
Jacobian, $q_j.weight$ is the quadrature weight of $q_j$, and
$J\left(q_j\right)$ is the element Jacobian at point $q_j$.
\begin{algorithm}[h!]
\begin{algorithmic}[100]
\FORALL{cells $i \in G$}
\FORALL{quadrature points $q_i \in i$}
\FOR{$\alpha=1:S$}
\FORALL{cells $j \in G$}
\FORALL{quadrature points $q_j \in j$}
\FOR{$\beta=1:S$}
\STATE{$\mathbf{U}  \leftarrow \mathbf{LandauTensor}\left(q_i.r,q_i.z,q_j.r,q_j.z\right)$  } 
\STATE{$\mathbf{K} \leftarrow \hat{\nu}_{\alpha\beta}\frac{m_o}{m_{\alpha}}\frac{m_o}{m_{\beta}} \mathbf{U} \cdot \nabla f_\beta\left( q_j \right) w_{q_j} $}
\STATE{$\mathbf{D} \leftarrow - \hat{\nu}_{\alpha\beta}\frac{m_o}{m_{\alpha}}\frac{m_o}{m_{\alpha}} \mathbf{U} f_\beta\left( q_j \right) w_{q_j} $}
\STATE{$\mathbf{C} \leftarrow FiniteElementAssemble\left(\mathbf{C}, w_{q_i} , \mathbf{K},  \mathbf{D} \right)$}
\ENDFOR
\ENDFOR
\ENDFOR
\ENDFOR
\ENDFOR
\ENDFOR
\end{algorithmic} 
\caption{Simple algorithm to compute Landau Jacobian $\mathbf{C}$ with state $f$}
\label{fig:algo0}
\end{algorithm}

The Landau tensor $\mathbf{U}$ in
(\ref{eq:fric_coef},\ref{eq:diff_coef}) is computed, or read from memory, in the
inner loop.  
A vector $\bm{K} = \mathbf{U} \cdot \nabla f_{q_j} w_{q_j}$
and a tensor $\mathbf{D} = \mathbf{U}f_{q_j} w_{q_j}$ are
accumulated in the inner loop.  With $S$ species, the accumulation of
$\bm{K}$ and $\mathbf{D}$ requires $6S$ words.  These
accumulated values are transformed in a standard finite element
process from the reference to the real element geometry and assembled
with finite element shape functions into the element matrix.
The six loops of Algorithm \ref{fig:algo0} can be processed in any
order, and blocked, giving different data access patterns, which is
critical in optimizing performance.

The first two issues that we address in the design of the Landau
solver are 1) whether to precompute the Landau tensors or compute them
as needed and 2) whether to use a single mesh with multiple degrees of
freedom per vertex or use a separate mesh for each species.


\subsubsection{To Precompute the Landau Tensor or Not to Precompute}
\label{sec:issues}

The Landau tensor is only a function of mesh geometry and can be
computed and stored for each mesh configuration.  The cost of
computing the Landau tensor is amortized by the number of nonlinear
solver iterations and the number of time steps that the mesh is used
for.  The computation of the tensor can be expensive, especially in the axisymmetric case which involves two different tensors and evaluation of elliptic integrals (see Appendix \cite{Hirvijoki2016}), requiring approximately 165 floating
point operations (flops) as measured by both the Intel Software
Development Emulator (SDE) and code analysis, including four logarithm
and square roots. Storing two such tensors requires eight words of storage,
64 bytes with double precision words.  There are \Order{N^2} unique
mesh ($i,j$) pairs for which the tensors are computed or stored, which is too much to fit in a
cache of any foreseeable machine with any reasonable degree of
accuracy (e.g., 64 megabytes with $N=10^3$).  The decision to
precompute or compute as needed depends on several factors.

A simple analysis on KNL suggest that both approaches are viable but
that trends in hardware works in favor of the compute as needed
approach.  Assuming the equivalent of 200 ordinary flops per axisymmetric Landau
tensor pair calculation and 64 bytes of data, the flop to byte ratio
is about three.  KNL has a theoretical peak floating point capacity of
about $2.6 \times 10^{12}$ flops/second and around $400 \times 10^9$
bytes/second on package memory bandwidth capacity, as measured by
STREAMS, or a flop to byte ratio of about six.  This simple analysis
suggests that precomputing would be two times slower, however, we
achieve about 20 \% of theoretical peak flop rate and, thus, a
precomputed implementation would need only to achieve about 40\% of
STREAMS bandwidth to match the run time of each kernel evaluation,
which one would expect is achievable.  This analysis suggests that
either method could be effective on KNL, but the spread between flop
capacity, in the form of more vector lanes and more hardware resources
per lane, and memory bandwidth capacity is anticipated to increase in
the future, which will benefit the compute on demand approach.

The kernel in Algorithms \ref{fig:algo0} requires $3S+1$ words from
memory per kernel evaluation for the weight $w_{q_j}$, the value
$f_\beta\left( q_j \right) $ and gradient of $f_\beta\left( q_j
\right)$ for each species.  The compute on demand approach also
requires the coordinate.  This is \Order{N} data, which has the
potential to fit in cache, for example, with $N=10^3$ and two species
this data would be about 64 kilobytes, plus lower order data, per
thread.  Eight threads per tile should fit in the 1Mb shared L2 cache
on KNL.

\subsubsection{Single and Multiple Meshes}
\label{sec:singlemesh}

We use a single mesh, independently adapted for each species, with $S$
degrees of freedom per vertex, however one can use multiple meshes or
a mesh for each species.  Observe that the integrals in
(\ref{eq:fric_coef},\ref{eq:diff_coef}) are decoupled from the outer integral in
(\ref{eq:K},\ref{eq:D}).  In theory, one can use a separate grid, or
different quadrature or even a different discretization, for each
species.  One could even use a different mesh for the inner and outer
integral in (\ref{eq:K},\ref{eq:D}).  An advantage of using a single mesh
is that the two loops over species in (\ref{eq:fullC}) can be
processed after the Landau tensor $\mathbf{U}$
is computed, and hence this tensor can be reused $S^2$ times.
However, if all of the species have ``orthogonal" optimal meshes, that
is each quadrature point only has significant information for one
species, which is a good assumption for ions and electrons because of
their disparate masses, then a single mesh requires about as many
vertices as the sum of each of the putative multiple meshes.  With the
model of orthogonal optimal meshes and kernel dominated computation or
communication, and with $N_\alpha$ quadrature points for each species
$\alpha$, the complexity of a Landau solve is $\mathcal{O}
\left(\sum_{\alpha=1}^S N_\alpha \right)^2$ for both the single and
multiple mesh approach.

With multiple ion species the orthogonal mesh assumption would be less
accurate, because (small) ions have similar optimal meshes, which
benefits the single mesh approach.  The Jacobian matrix for the single
mesh approach has about $S$ times more non-zeros, which is not
important if the total cost is dominated by the Landau kernel.  It is
likely that with further optimization of the Landau kernel, the next
generations of hardware, and algorithmic improvements for the inner
integral, that the inner integral costs will be reduced relative to
the rest of the solver costs, which would benefit the multiple mesh
approach.  The single mesh method has larger accumulation register
demands and larger element matrices, which pressures the memory system
more and is advantageous for the multiple mesh approach.  The result
of the increased register pressure can be seen in the decreases flop
rates in Table \ref{tab:species} with the increase in the number of
species.

Another potential advantage of the single grid method is that the
extra degrees of freedom, in for instance the ions in the range of one
electron thermal radius, might be beneficial.  The large scale
separation between ions and electrons means that small relative errors
in the ion distribution can be large relative to the electron
distribution.  Ions and electrons have about equal and opposite
charge, cancellation could lead to large relative errors in the total
charge density.
The accuracy of collisions with fast electrons in the tail of the ion
distribution could benefit from extra resolution.  A more thorough
understanding of the accuracy issues would be required to fully
address the question of using a single or multiple meshes.

\subsection{Our Algorithm}
\label{sec:algo}

For demonstration purposes, we focus on implementing the axially
symmetric version using cylindrical velocity coordinates $\bm{x}=(r,\theta,z)$. Under axial symmetry the distribution function is independent of the angular velocity coordinate ($\partial_{\theta}f=0$) and the evaluation of the vector $\bm{K}$ and the tensor $\mathbf{D}$ requires two different Landau tensors $\mathbf{U_K}$ and $\mathbf{U_D}$ respectively (see Appendix \cite{Hirvijoki2016}). We choose to compute the
required Landau tensors as needed and use a single mesh with a
degree-of-freedom for each species on each vertex.  We fuse the two
inner loops over cells and quadrature points, inline the function call
of, and within, the Landau tensor function.  Algorithm \ref{fig:algos}
shows the initialization of the vectors $r$, $z$, $w$, $f$, and the
two gradient vectors $df[1]$ and $df[2]$, with $|G|$ cells in the set
$G$, $S$ species, and weights $w_{q_i}$ at each quadrature point $i$.
Each quadrature point $q_i$ is located at a 2D coordinate ($q_i.r$,
$q_i.z$).
\begin{algorithm}[h!]
\begin{algorithmic}[1]
\FORALL{cells $i \in G$}
\FORALL{quadrature points $q_i \in i$}
\STATE{$r.append(q_i.r)$}
\STATE{$z.append(q_i.z)$}
\STATE{$w.append(w_{q_i})$}
\FOR{$\alpha=1:S$}
\STATE{$f[\alpha].append(f_\alpha(q_i) $)}
\STATE{$df[1][\alpha].append(\nabla f_\alpha(q_i)[1] $)}
\STATE{$df[2][\alpha].append(\nabla f_\alpha(q_i)[2] $)}
\ENDFOR
\ENDFOR
\ENDFOR
\end{algorithmic} 
\caption{Initialization of vectors $r$, $z$, $w$, $f$, and $df$ with state $f$}
\label{fig:algos}
\end{algorithm}

Algorithm \ref{fig:algo2} shows the algorithm for the construction of
the Landau collision integral Jacobian.
\begin{algorithm}[h!]
\begin{algorithmic}[1]
\FORALL{cells $i \in G$}
\STATE{$\mathbf{ElemMat} \leftarrow 0$}
\FORALL{quadrature points $q_i \in i$}
\STATE{$\mathbf{K} \leftarrow 0$}
\STATE{$\mathbf{D} \leftarrow 0$}
\STATE{$w_i \leftarrow q_i.weight \cdot \left| J\left(q_i\right) \right| \cdot q_i.r $}
\STATE{$N \leftarrow Nq \cdot |G|$}
\FOR[Vectorized loop]{$n=1:N$}
\STATE{$\left[\mathbf{U_K}, \mathbf{U_D} \right]  \leftarrow \mathbf{LandauTensor}\left(q_i.r,q_i.z,r[n],z[n]\right)$  } 
\FOR{$\alpha=1:S$}
\FOR{$\beta=1:S$}
\STATE{$\mathbf{K}\left[{\alpha}\right]  \leftarrow \mathbf{K}\left[{\alpha}\right]  + \hat{\nu}_{\alpha\beta}\frac{m_o}{m_{\alpha}}\frac{m_o}{m_{\beta}} \mathbf{U_K} \cdot df[:][\beta][n] w[n]$}
\STATE{$\mathbf{D}\left[{\alpha}\right]  \leftarrow \mathbf{D}\left[{\alpha}\right]  - \hat{\nu}_{\alpha\beta}\frac{m_o}{m_{\alpha}}\frac{m_o}{m_{\alpha}} \mathbf{U_D} f[\beta][n] w[n]$}
\ENDFOR
\ENDFOR
\ENDFOR
\FOR{$\alpha=1:S$}
\STATE{$\mathbf{G2}\left[{\alpha}\right]  \leftarrow \mathbf{J}\left(q_i\right)^{-1} \mathbf{K}\left[{\alpha}\right]  w_i$} \COMMENT{transform point integral to global space}
\STATE{$\mathbf{G3}\left[{\alpha}\right]  \leftarrow \mathbf{J}\left(q_i\right)^{-1} \mathbf{D}\left[{\alpha}\right]  \mathbf{J}\left(q_i\right)^{-1} w_i$}
\ENDFOR
\STATE{} \COMMENT{Project point value to vertices of cell $i$}
\STATE{$\mathbf{ElemMat} \leftarrow Transform\&Assemble\left(\mathbf{ElemMat}, \mathbf{G2}, \mathbf{G3},  \mathbf{B}\left(q_i\right) \right)$}
\ENDFOR
\STATE{} \COMMENT{Sum element matrix into global Jacobian}
\STATE{$\mathbf{C} \leftarrow GlobalAssemble\left(\mathbf{C}, i, \mathbf{ElemMat} \right)$}
\ENDFOR
\end{algorithmic} 
\caption{Algorithm to compute $\mathbf{C}$ with $r$, $z$, $w$, $f$, and $df$ from Algorithm \ref{fig:algos} }
\label{fig:algo2}
\end{algorithm}
This algorithm is designed to minimized data movement by computing the
Landau tensors as needed and exploits a single mesh by lifting the
kernel outside of the two inner loops over species.

\pagebreak

\section{Numerical Methods and Implementation}
\label{sec:numer}

We implement the Landau solver with the PETSc numerical library
\cite{petsc-web-page,petsc-user-ref}.  PETSc provides finite element
(FE) and finite volume discretization support, mesh management,
interfaces to several third party mesh generators, fast multigrid
solvers, interfaces to several third party direct solvers, AMR
capabilities among other numerical methods.  We adapt nonconforming
tensor product meshes using the third party {\it p4est} library
\cite{Stadler1033,DBLP:journals/siamsc/IsaacBWG15,Rudi:2015:EIS:2807591.2807675},
and unstructured conforming simplex meshes with PETSc's native AMR
capabilities \cite{BarralKnepleyLangePiggottGorman2016}.  Our
experiments use bi-quadratic (Q2) elements with {\it p4est}
adaptivity, with the PETSc's Plex mesh management framework.

The computational domain is $\Omega=\{(r,z) \mid 0 \le r \le L, -L\le
z \le L\}$, with $L=2$.  We use Neumann boundary conditions and
shifted Maxwellian initial distribution functions, for each species,
of the form
\begin{align*}
\label{eq:max}
f_{\alpha}(\bm{x},t=0)=\theta \frac{1}{2}\left(\pi\sigma_{\alpha}^2\right)^{-3/2} \exp\left(-\frac{r^2+(z-s_{\alpha})^2}{\sigma_{\alpha}^2}\right),
\end{align*}
where $\sigma_{\alpha}=2T_{\alpha}m_o/m_{\alpha}$, $s_i = 0$, $s_e =
-1$, $T_{\alpha}$ is temperature, and $\theta$ is a scaling factor
used to maintain quasi-neutrality.

We solve the boundary value problem $$\frac{\partial
  f_{\alpha}}{\partial t}\left({\bf v}, t \right) -
\mathbf{C}_{\alpha}[f] f = 0$$ 
in axisymmetric coordinates, with standard FE methods and time
integrators.
A Newton nonlinear solver with the SuperLU direct liner solver is used
at each time stage or step \cite{Li:2005:OSA:1089014.1089017}.  These
experiments use a Crank-Nicolson time integrator.

A global kinetic model would include a 3D spatial component and the 3V version of this
solver would be used at each cell in either a particle method
\cite{Hager2016}, or a grid based kinetic method \cite{Candy2014}.
Our numerical experiments use up to 272 Message Passing Interface
(MPI) processes on one KNL socket, redundantly solving the problem, to
include some of the memory contention of a full 5D code.  The timing
experiments run one time step with one Newton iteration, which results
in the Landau operator being called twice (one more than the number of
Newton iterations), and with one linear solve (one per Newton
iteration).  The time for this step is reported, which does not
include the AMR mesh construction.  We observe a significant
variability in times with 272 processes and have run each test several
times in several sessions, in both batch and interactive sessions, and
the report the fastest observed time.  We report the maximum time from any processor and see about a
10\% ratio between the maximum and the minimum time of any process
with large process counts.

\subsection{Overview of application and numerical method}
\label{sec:overview}

\begin{figure*}[!ht]
\includegraphics[width=.73\linewidth]{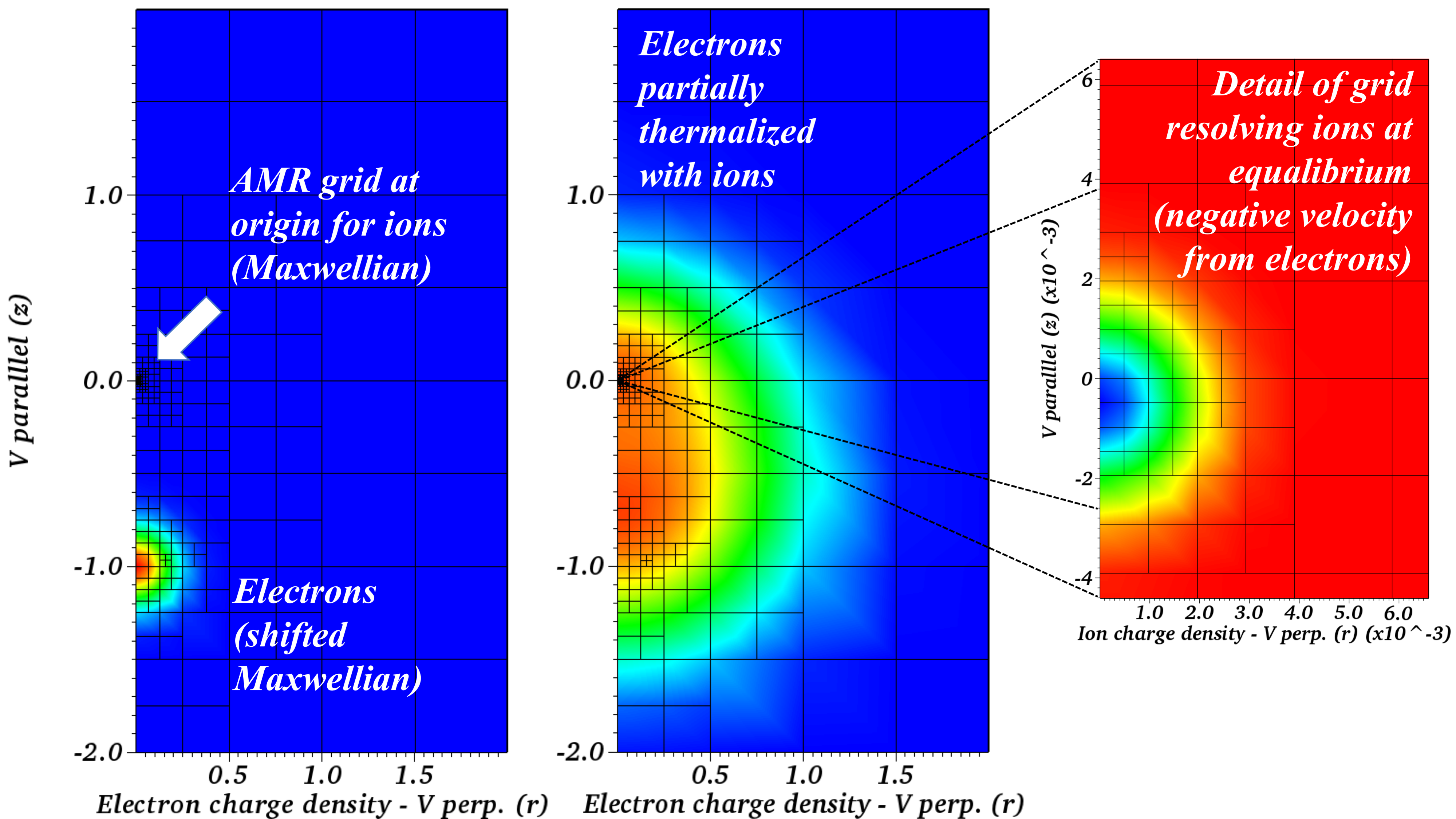}
\includegraphics[width=.26\linewidth]{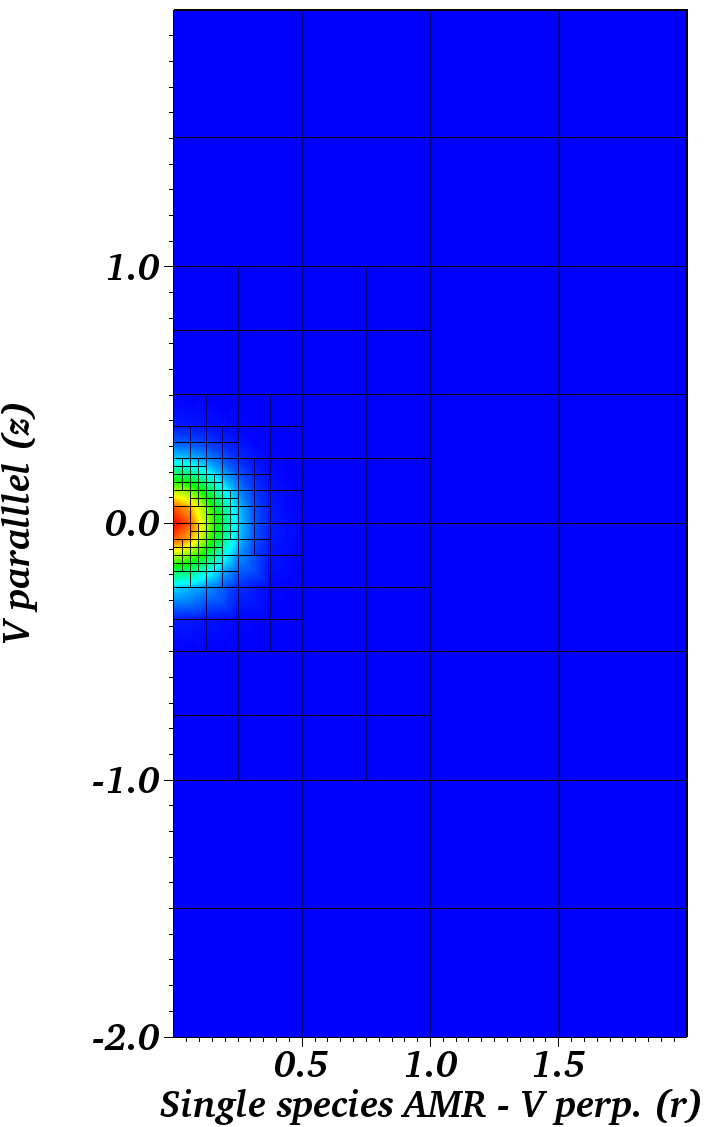}
\caption{Charge density with initial Maxwellian distribution functions relaxing towards equilibrium, initial electron distribution (left), partially thermalized electrons (center, left), detail of the ion distribution (center, right), electron only adapted mesh (right)}
\label{fig:evolution}
\end{figure*}

To illustrate the capabilities and behavior of the solver, we run the
code to near equilibrium, initializing electrons with a shifted
Maxwellian distribution hitting a stationary single proton ion
population with a Maxwellian distribution.  We use a realistic mass
ratio of $\frac{m_i}{m_e}=1836.5$ and temperatures of $T_e = 0.02$ and
$T_i = 0.002$.  Figure \ref{fig:evolution} (left) shows the initial
electron distribution with the ion grid at the origin, a partially
thermalized electron distribution (center, left), and Maxwellian ion
distribution near equilibrium (center, right).  The ion distribution
has been shifted from the origin by collision with the electrons.
The ions are resolved with AMR at the origin and have a near
Maxwellian distribution.  Note, Figure \ref{fig:evolution} uses linear
interpolation from the four corners of each quadrilateral, whereas the
numerics use bi-quadratic interpolation with nine vertices per
quadrilateral, which results in inaccuracy and asymmetries in the
plots not present in the numerics.

\subsection{Optimization and Performance}
\label{sec:opt}

Most of the work in the Landau solver is in the inner integral of
(\ref{eq:K},\ref{eq:D}) (lines 8-16 in Algorithm \ref{fig:algo2}).  This
kernel is vectorized with Intel AVX-512 intrinsics.  The Landau
tensors calculation includes two logarithms, a square root, a power,
seven divides, about 85 multiplies and 165 total flops.  The power is
converted to a inverse square root, an intermediate divide is reused,
resulting in five divides, two logarithms, a square root, and an
inverse square root.  The KNL sockets used for this study is equipped
with 34 tiles, each with a 1 megabyte shared L2 cache and two cores,
each core has two 8 lane vector units that can issue one fused
multiply add (FMA) per cycle per lane.  Each core has four hardware
threads, for a total of 272 threads per socket.  The KNL clock rate is
1.4 GHz, but is clocked down to 1.2 GHz in AVX-512 code segments.
This results in a theoretical flop peak rate of $2.6 \times 10^{12}$
flops/second.  The peak flop rate that can be achieved with this
solver is reduced because the kernel is not entirely composed of FMAs
and the four logarithms and square roots require considerably more
than one cycle each.

The performance data in this section uses a simplified version the
test problem, a grid adapted for electrons only with 176 cells and
1,584 quadrature points, a mass ratio of $\frac{m_i}{m_e}=1$ and $T_e
= T_i = 0.2$, and no Maxwellian shifts ($s_i = s_e = 0$), as shown in
Figure \ref{fig:evolution} (right).

\subsection*{Performance overview}
\label{sec:perf1}

The major code segments have been instrumented with PETSc timers.
Table \ref{tab:overview} show the maximum time from any process for major components of the Landau operator, the total Landau operator, and the
linear solver.
\begin{table}[!ht]
\begin{center}
\begin{tabular}{|l||l|l|}
\hline
Component (times called) & Time (maximum) & \% of total \\
\hline
Landau initial vector data setup, Algorithm \ref{fig:algos}    &  0.019 &  2 \\
 Landau kernel with AVX512 intrinsics &  0.533 & 66  \\
Landau FE transforms \& assemble &  0.030 & 4  \\
Landau FE global matrix assembly & 0.072 & 9  \\
 \hline
 Landau operator total (2) &  0.682 &  85  \\
 Linear solver (1) & 0.12 & 15  \\
 \hline
 Total time step time (1) & 0.803 & 100 \\
 \hline
 \end{tabular}
\end{center}
\caption{Major component times (maximum of any process) from one time step with two species, double precision and 272 processes}
\label{tab:overview}
\end{table}

This data is from the double precision solve with 272 processes and
two species in Table \ref{tab:single}.  This data shows that the
Landau kernel, though vectorized, is still responsible for most of the run time.

\subsection*{Performance and complexity analysis}
\label{sec:model}

There are two types of work in the kernel: 1) computing the two Landau
tensors and 2) the accumulation of $\bm{K} $ vector and
$\mathbf{D}$ tensor.  The accumulation requires $20S^2$ flops
(lines 12-13 in Algorithm \ref{fig:algo2}).  Instrumenting this inner
loop would be invasive, but we can infer the percentage of time and
work in these two parts with a complexity model and global
measurements.  Assume both the time and work cost of the entire solver
are of the form $C = aS^0 + bS^1 + cS^2$.  The solve time and flop
counts with $S=1,2,3$, shown in Table \ref{tab:species}, generate
right hand sides for system of three equations and three unknows $a, b$, and $c$, which are the time spent, or work, in each
of the three types of components.
\begin{table}[!ht]
\begin{center}
\begin{tabular}{|c||c|c|c|c|}
\hline
 \# Species  & 1 proc. &  272 proc. & Gflops 1 proc. & Gflops/sec. (\% of peak) \\
\hline
1  &  0.21 &  0.47 & 1.01 & 572 (22) \\
2  &  0.28 &  0.79 & 1.34 & 455 (18) \\
3  &  0.38 &  1.38 & 1.88 & 370 (14) \\
\hline
\end{tabular}
\end{center}
\caption{Time (seconds) with 1 and 272 processes on one KNL socket; flop counts from Intel SDE, and flop rates}
\label{tab:species}
\end{table}
The Landau tensor cost is formally independent of the number of
species, and the work in the accumulation has $\Order{S^2}$ work
complexity.  Most of the rest of the costs, given a mesh, order of
elements, etc., has $\Order{S^1}$ complexity.  Table \ref{tab:model}
shows the percentage of time and work in each component, infered from
the data in Table \ref{tab:species}.
\begin{table}[!ht]
\begin{center}
\begin{tabular}{|c||c|c|}
\hline
Work type  & Flops  \% & Time \% \\
\hline
$S^0$ (Landau tensors) &  88 & 81 \\
$S^1$ (non-kernel work)  &  1.5 & 12 \\
$S^2$ (accumulation)  &  10.5 & 7 \\
\hline
\end{tabular}
\end{center}
\caption{Percentage of flops and time in species independent work, work linear in $S$, and work quadratic in $S$, with two species, double precision, and one process}
\label{tab:model}
\end{table}
This analysis shows that about $98\%$ of the work, and about $88\%$ of
the time, is in the kernel with one process.  The measurements of the
kernel time in Table \ref{tab:overview} is $66\%$ with 272 processes.
This discrepancy is probably due to performance noise and memory
contention in the 272 process timings.  The non-kernel time percentage
(12) increases by a factor of about eight from the flop percentage
(1.5), which reflects the eight vector lanes of the KNL vector unit.

\subsection*{Memory performance}

Our experiments are trivially parallel, but memory contention results
in performance degradation as more processes are used on a socket.
KNL's architecture allows for twice as many vector lanes with single
(32 bit) versus double (64 bit) precision and can thus run, in theory
twice as fast in single precision.  This section investigates memory
issues with a weak speedup study in single and double precision.
Table \ref{tab:single} shows timing data with increasing number of
processes on a single KNL socket, with single and double precision.
\begin{table}[!ht]
\begin{center}
\begin{tabular}{|c||c|c|c|c|c|}
\hline
 Processes  & 272 & 136 & 68 & 34 & 1\\
\hline
Single  & 0.51   & 0.29 & 0.18 &  0.17 & 0.17 \\
Double & 0.80  &  0.46 & 0.30 &  0.28 & 0.28 \\
\hline
\end{tabular}
\end{center}
\caption{Weak speedup time (seconds) with single and double precision}
\label{tab:single}
\end{table}
This data shows that we are achieving about $80\%$ of the perfect
factor of two speedup with single precision.

A KNL socket has 136 vector units.  One might expect that using more
processes than 136 would not be useful, however, the kernel has serial
dependancies that result in bubbles in the pipeline, especially in the
ninth and tenth order polynomial evaluations in the elliptic
integrals.  These holes can be filled by interleaving a second process
in the second hardware thread.  We do see about a 15\% increase in
total performance from the added parallelism of using 272 processes.
This data shows that the flop rate per process decreases by a factor
of about 3 in going from one process per tile to two processes per
vector unit, or eight processes per tile.  There is no difference
between the one process and 34 process runs, suggesting that the 34
processes are indeed placed with one per tile.  There is little
degradation going from one to two processes per tile, suggesting that
the problem still fits in the L2 cache.  The degradation from one to
two processes per core is from L1 cache misses because four processes
should be able to fit into the 1 megabyte L2 cache.  Analysis of
memory complexity and the time data going from one to two processes
per vector unit suggest that the full 272 process runs fit in the L2
cache, or nearly so.

\subsection{Verification}

\begin{figure*}[!ht]
\subfloat[Thermal flux (top), and flux differences\label{sbfig:1}]{%
  \includegraphics[width=.5\linewidth]{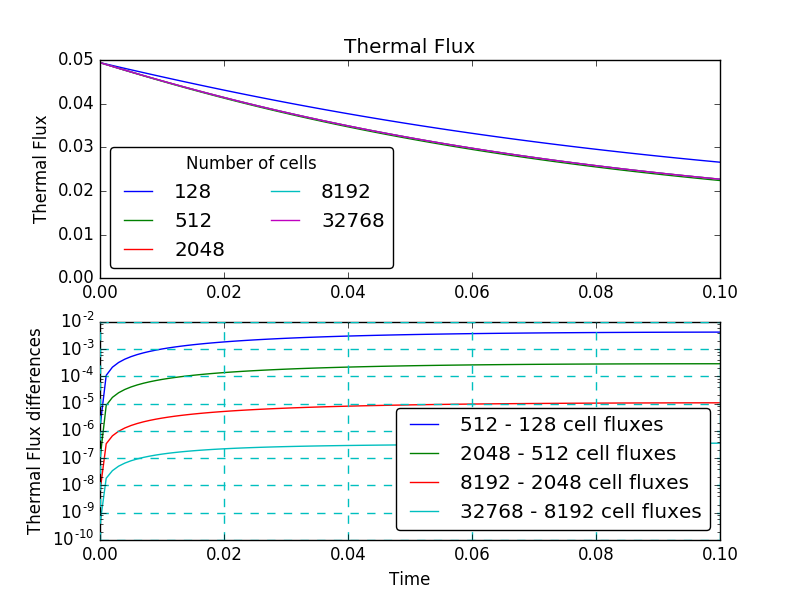}%
}\hfill
\subfloat[Error vs. cell count\label{sbfig:2}]{%
  \includegraphics[width=.5\linewidth]{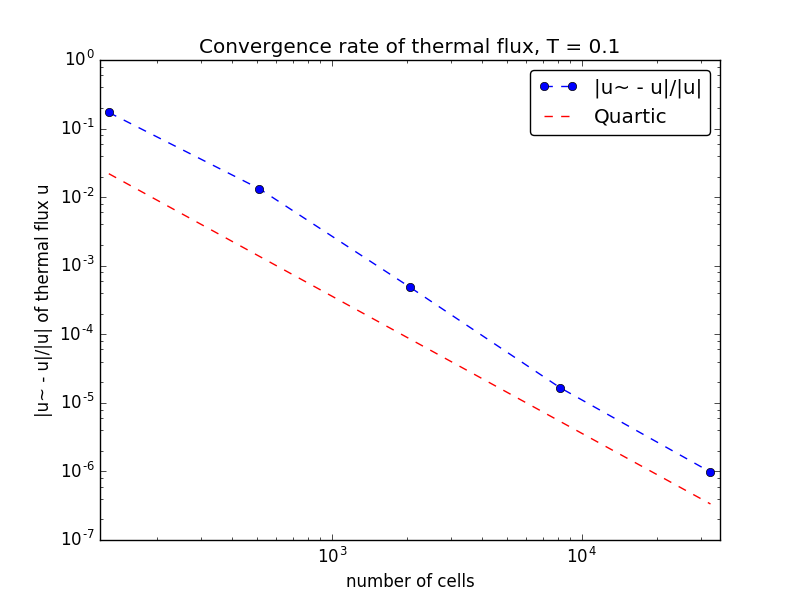}%
}
\caption{ (a) Therm flux over time (top, left), flux differences in grid sequence (bottom, left), quartic convergence rate (right)}
\label{fig:conv}
\end{figure*}

Consider a convergence study with Cartesian grids of the third moment,
thermal flux, to verify the expected order of convergence.  An
analytical flux for this problem does not exist.  Richardson
extrapolation is used to construct an approximate exact flux.  The
mass ratio is $4$, $T_e = 0.2$ and $T_i = 0.02$.  The flux history,
with a series of refined grids starting with a $8 \times 16$ grid, is
shown in Figure \ref{fig:conv} (top, left).  Figure \ref{fig:conv}
also shows the differences between fluxes on successive grids, and the
error convergence.

We can see from this data that we achieve fourth order convergence.

\section{Closure}

We have implemented a high order accurate finite element
implementation of the Landau collision integral with adaptive mesh
refinement in the PETSc library using AVX-512 intrinsics for the
Second Generation Intel Xeon Phi, Knights Landing, processor.  We have
developed a memory centric algorithm for emerging
architectures that is amenable to vector processing.  We have achieved up to 22\% of
the theoretical peak flop rate of KNL and analyzed the performance
characteristics of the algorithm with respect to process memory
contention, single and double precision, and the results of
vectorization.
We have verified fourth order accuracy with a
bi-quadratic, Q2, finite element discretization.
Future work includes, building models for runaway electrons
in tokamak plasmas with this kernel \cite{Chang2015,Decker16,Nilsson15,Boozer2016}, 
and building up complete kinetic models (6D AMR) that also preserve the geometric 
structure of the governing equations of fusion plasmas \cite{2016arXiv161206734M}.

\section{Acknowledgments}

We are grateful for discussions and insights from Intel engineer Vamsi
Sripathi.  This work has benefited from many discussions with Sam
Williams.  This work was partially funded from
the Intel Parallel Computing Center program.  This research used
resources of the National Energy Research Scientific Computing Center,
a DOE Office of Science User Facility supported by the Office of
Science of the U.S. Department of Energy under Contract
No. DE-AC02-05CH11231.

\bibliographystyle{siamplain}
\bibliography{refs}

\end{document}